\documentclass{iopart}
\usepackage{amsbsy}
\usepackage{amssymb}
\usepackage{esint}
\usepackage{multicol}
\makeatletter
\usepackage{iopams}
\usepackage{setstack}
\usepackage{url}
\usepackage{graphicx}
\usepackage{dcolumn}
\usepackage{bm}
\usepackage{pgfplotstable} 

\def\lsim{\mathrel{\rlap{\lower4pt\hbox{\hskip1pt$\sim$}}
\raise1pt\hbox{$<$}}}
\def\gsim{\mathrel{\rlap{\lower4pt\hbox{\hskip1pt$\sim$}}
\raise1pt\hbox{$>$}}}

\def\draftmark {
}

\newcommand{\Note}[1]{[$\blacktriangleright$~\textbf{#1}~$\blacktriangleleft$]}

\begin{document}

\draftmark

\title[Limiting alternative theories of gravity using gravitational waves]{Limiting alternative theories of gravity using gravitational wave observations across the spectrum}
\author{Jeffrey S. Hazboun,$^1$ Manuel Pichardo Marcano,$^2$ and Shane L.\ Larson$^3$}

\address{$^{1}$ Department of Physics, Utah State University, Logan, Utah 84322}

\address{$^{2}$ Department of Physics, Utah State University, Logan, Utah 84322}

\address{$^3$ Center for Interdisciplinary Exploration and Research in
   Astrophysics (CIERA) \& Department of Physics and Astronomy,
   Northwestern University, Evanston, IL 60208}
\ead{\mailto{jeffrey.hazboun@gmail.com}  \mailto{manuelmarcano22@yahoo.com}
\mailto{s.larson@northwestern.edu}}

\date{\today}

\begin{abstract}
The advent of gravitational wave astronomy provides new proving
grounds for testing theories of gravity. Recent work has
reinvigorated the study of bimetric theories of gravity and massive
gravity theories.  One of the most interesting predictions of these
theories, as well as some string theories, is the subluminal speed of propagating gravitational waves.
Multi-messenger astronomy provides a unique opportunity to put limits
on the difference (either positive or negative) between the propagation speed of electromagnetic and
gravitational waves from these sources.  This paper considers limits from
multi-messenger cases across the planned measurable spectrum: first, the limits from isolated pulsars based
on the current best limits from LIGO on gravitational wave emission,
second, the limits from ultra-compact binaries that will be visible
to a low-frequency space-based gravitational wave observatory like
LISA, and third, limits from super massive black hole binaries using pulsar timing arrays. The required phase comparison between the electromagnetic signal
and the gravitational wave signal is derived and, assuming a null
result in that comparison, the current bounds on emission are used to
place limits on alternative theories that exhibit propagation delays.
Observations of the pulsars in the most sensitive range of LIGO could
put an upper limit on the graviton mass as low as
$10^{-38}\frac{eV}{c^{2}}$ and an upper limit on the fractional
difference between the gravitational wave and electromagnetic wave
speeds as low as $10^{-9}$. This paper shows results from the initial LIGO limit catalog of known pulsars. The bounds are stronger for binaries. A LISA-like mission bounds $m_{g}<10^{-40}\frac{eV}{c^{2}}$ and $\delta<10^{-12}$. A PTA source gives even better bounds of $m_{g}<10^{-45}\frac{eV}{c^{2}}$ and $\delta<10^{-14}$.
\end{abstract}

\pacs{04.30.-w, 04.30.Tv, 04.80.Nn, 95.55.Ym}

\submitto{\CQG}

\maketitle

\section{Introduction}\label{sec.intro}
One of the first dynamical results derived from general relativity was
the theoretical prediction of gravitational wave solutions to the
field equations by Einstein \cite{einsteinGW1918}.  Initially
discounted by Einstein as undetectable, owing to their extremely small
coupling to physical detectors, attempts to experimentally observe
these waves did not begin in earnest until the 1960s with the
development of resonant bar detectors.  In the time since those
pioneering efforts, technology has expanded the field of gravitational
wave astronomy dramatically with the advent of broadband detection
schemes across the entire gravitational wave spectrum.  The very
low-frequency regime ($f_{gw} \sim 10^{-9}$ Hz) is currently being
explored using pulsar timing techniques \cite{iptaStatus2010}, and in
the next decade the low-frequency band ($f_{gw} \sim 10^{-3}$ Hz) will
be probed by space-based gravitational wave interferometers, the
archetype of which has been the LISA design concept \cite{lisa2003}.
The high-frequency band ($50 {\rm Hz} \lesssim f_{gw} \lesssim 1000$
Hz) is being covered by a globe-girdling network of kilometer-scale
gravitational wave interferometers on the ground, including the
European GEO-600 \cite{geoStatus2008} and Virgo \cite{virgoStatus2010}
detectors, and a pair of American detectors known as LIGO
\cite{ligoStatus2010}.  The design sensitivity of these detectors
will, within the next decade, reach a level where regular detection of
gravitational wave sources will become a common occurrence, giving
birth to a new branch of observational science, gravitational wave
astronomy \cite{ligoObservingPlan2013}.

The advent of gravitational wave astronomy promises to dramatically
expand our ability to probe astrophysical systems, since gravitational
waves encode detailed information about the bulk distribution and
motion of matter, information that is highly complementary to the
usual information carried by photons.  Direct detection of
gravitational waves promises to provide accurate luminosity distances
to binary systems \cite{schutz1986}, probes of the engines for
gamma-ray bursts \cite{shortGRB}, information about the shape
\cite{ligo116pulsarsS5} and equation of state of neutron stars
\cite{vallisneriEOS,ligoEOS}, and many other astrophysical phenomena.
These astrophysical results can be amplified when combined with
associated electromagnetic observations.  \textit{Multi-messenger}
observations propose to synthesize a coherent picture of an
astrophysical system by using simultaneous gravitational wave (GW) and
electromagnetic (EM) observations to constrain the physical state of
the source.

Another important application of multi-messenger observations is
testing gravitational theory.  There have been many proposals whereby
gravitational wave observations could be used to test the precepts of
general relativity using a variety of astrophysical systems observed
by ground-based and space-based gravitational wave detectors
\cite{1998PhRvD..57.2061W,2009PhRvD..80d4002S,2009CQGra..26o5002A}.
Of particular interest in this paper are \textit{propagation tests},
where the emission and arrival of gravitational waves are compared
directly to electromagnetic waves.  Such comparisons allow a variety
of gravitational theories to be tested, including massive graviton
theories, 
and bimetric theories \cite{1998PhRvD..57.2061W,Will:1993ns}.  Both
will be examined in detail here.

The outline of the paper is as follows.  In Section 2 we briefly
outline the premise of alternative theories of gravity, including
bimetric theories of gravity and massive graviton theories.  In
section 3 we outline types of astrophysical systems that can be used
for establishing bounds based on propagation tests, and outline the
premise of the signal comparison; the sources of interest include
low-frequency, ultra-compact binaries (a prospective source for
space-based detectors), and isolated, continuous wave sources such as
pulsars (a prospective source for ground-based detectors).  Section 4
and 5 outline the bounds that can be expected based on the physical
parameters of detectable systems, and section 6 concludes with
discussion and speculation on future work.


\section{Alternative theories of gravity}\label{sec.Altgrav}

The classical tests of general relativity proposed by Einstein were
geared toward validating the \textit{predictions} of relativity, and
many considerations today are still focused on this goal, mounting up
the decimal places of agreement to higher and higher precision.  As
our sophistication with gravitational theory has grown, it has become
more common to consider \textit{alternative} theories of gravity
which may predict different behavior for physical systems than general
relativity.  The predicted behaviors may be dramatically different
from general relativity in strongly gravitating systems, but may
produce measurable differences, as well, in systems which only deviate
slightly from Newtonian gravity.  Gravitational waves offer an
important tool for probing the existence of alternative theories,
since they often predict different polarization structures for the
gravitational waves \cite{2012PhRvD..86b2004C}, or, as considered in
this work, different propagation speeds $v_{gw}$ for the waves
(general relativity predicts that $v_{gw} = c$).

One example are \textit{bimetric theories} of gravity, where the
metric that characterizes the pathways followed by light and
gravitational waves may be different \cite{Eardley:1973p2936}.  These
theories, first examined by Rosen and others in the 1970's, have seen
a recent resurgence since 2011 when Hassan and Rosen
\cite{Hassan:2011p3350} showed that there is a connection between
massive gravity theories and bimetric theories.  Massive gravity
theories, first proposed by Fierz and Pauli \cite{Fierz:1939ix}, have
had a very dynamic history.  They were discounted in the 1960's
because of the vDVZ discontinuity (van Dam, Veltman, Zhakarov)
\cite{vanDam:1970vg, Zakharov:1970cc}, which was later found to be a
gauge effect solved using the Vainshtein mechanism
\cite{Vainshtein:1972sx}.  Later the existence of Boulware-Deser
ghosts (negative norm states) \cite{Boulware:1973my} was found in
these theories.  In 2011 the deRham, Gabadadze, Tolley model
\cite{deRham:2011rn} showed that a large class of massive gravity
theories did not exhibit pathological behavior.  In the same year
Hassan and Rosen showed that this class of theories is free of ghosts.
Rosen and Hassan \cite{Hassan:2011p3350} later showed that bimetric
theories can be derived from these new massive gravity theories, and
so the theories have been connected.  The theoretical debate is far
from over as issues have been found concerning faster than c speeds and
acausality \cite{Izumi:2013poa, Deser:2013eua}.  Various versions of
string theory have also been shown to admit an effective mass for the
graviton \cite{Kostelecky:1990pe}.  The purpose of the current work is
to look at experimental bounds on all of these theories using
gravitational wave propagation speed.

In some classes of bimetric theories, one or more polarizations of the
gravitational waves propagate along null geodesics of a flat spacetime
described by the metric $\eta_{\mu\nu}$, \cite{Mohseni:2012p3424,
Eardley:1973p2936} whereas electromagnetic waves propagate on the null
geodesics of a ``physical metric'' $g_{\mu\nu}$.  If $|\eta_{\mu\nu}|$
and $|g_{\mu\nu}|$ are typical magnitudes of the metric elements, then
an experimental bound on the effect of the bimetric theory on
propagating signals is
\begin{equation}
	\delta > \frac{|g_{\mu\nu} - \eta_{\mu\nu}|}{|\eta_{\mu\nu}|}
	= \frac{|v_{gw} - v_{em}|}{c}\ ,
	\label{eqn.delta}
\end{equation}
where $v_{gw}$ is the propagation speed for gravitational waves and
$v_{em}$ is the propagation speed for electromagnetic waves.  Directly
measuring the propagation speed between simultaneous gravitational and
electromagnetic wave signals emitted by an astrophysical system yield
experimental values of $\delta$; as $\delta$ becomes smaller ($\delta
\rightarrow 0$ is the GR limit), bimetric theories become more
strongly constrained.  Sufficiently small values of $\delta$ will rule
out different bimetric theories. These observations are able to detect 
negative values of $\delta$ as well, and are therefore able to rule out 
superluminal propagation. The bimetric theory discussed in
\cite{Gumrukcuoglu:2012wt} lists $\delta=10^{-27}- 10^{-22}$ as the
range to look for interesting observational features, while some
massive graviton theories, able to explain the acceleration of the
expansion of the universe without dark energy, require the mass of the
graviton to be on the order of $10^{-49}\frac{eV}{c^{2}}$
\cite{Gong:2012ny,Gratia:2012wt}.  

\section{Astrophysical signals}\label{sec.astro}

In order to consider measurements of $\delta$ in Eq.\ \ref{eqn.delta},
astrophysical systems must be identified where electromagnetic signals
can be detected simultaneously with gravitational wave signals.  For
the purposes of making a comparison $|v_{gw} - v_{em}|$, long-lived
continuous sources are of the most interest.  Different astrophysical
systems will have different detectability on both the electromagnetic
and gravitational wave fronts.  Here we consider sources, in the
context of high-frequency (ground-based) gravitational wave sources,
low-frequency (space-based) gravitational wave sources,
and ultra low-frequency (PTA-based) gravitational wave sources.

\subsection{Electromagnetic Signals}\label{sub.emSignals}
In the high-frequency gravitational wave band, covered by ground-based
detectors like LIGO and Virgo, the prospective continuous wave source
of interest for this study are isolated pulsars.  The electromagnetic
signal from pulsars is a periodic pulse of electromagnetic energy
recurring at the spin frequency of the pulsar.  The detectability of a
given pulsar depends on its luminosity, distance, and the sensitivity
of the telescope being used to observe it.  There are some 2000 radio
pulsars known \cite{2005AJ....129.1993M}, though ongoing surveys are
continuing to search for and add to this number.  Future large scale
radio surveys, enabled by instrumentation such as the Square Kilometer
Array \cite{2013IAUS..291..337T} are expected to increase the size of
this catalog by an order of magnitude.  Currently the most distant
known pulsar (and the most luminous) is J1823-3021A, 27,000 light
years away in globular cluster NGC 6624.  The number of systems that
could be probed for the purposes of this work is roughly the Milky Way
space density of pulsars multiplied by a sphere whose radius is equal
to the most distant reach of our telescopes.  If the distance to PSR
J1823-3021A is characteristic of the maximum distance to which a
pulsar can be detected, then current technology would allow us to
probe a volume that covers roughly half the galaxy; future high
sensitivity surveys will expand this volume.

In the low-frequency gravitational wave band, covered by space-based
detectors similar to the classic LISA concept \cite{lisa2003}, such as
eLISA \cite{2013arXiv1305.5720C} and SGO \cite{SGO}, ultra-compact
binaries are a guaranteed source (for all mission concepts under
consideration).  These systems are short period binaries comprised of
two stellar remnants; white dwarfs are the most common, followed by
neutron stars and black holes.  Population synthesis models have
estimated that the population of these sources will number in the tens
of millions (e.g. \cite{2001AA...375..890N}).  The total population in
the galaxy will be so large that the signals will be overlapping and
unresolved at frequencies $f \lesssim 3$ mHz.  Studies have suggested
that thousands of individual systems will be individually resolvable
\cite{2007CQGra..24S.575C}\cite{2008CQGra..25k4037B}; a significant
subset of these are expected to be detectable by both gravitational
wave detectors and electromagnetic telescopes \cite{2013MNRAS.429.2361L}.

In the nano-hertz band of the gravitational wave spectrum, observed by
pulsar timing arrays, the sources are expected to be merging
supermassive black holes in the centers of galaxies.  These systems
arise from the mergers of galaxies.  The black holes initially find
themselves far apart, embedded in the dynamic cloud of stars from the
collision.  Over time they sink together toward the center of the next
galaxy via dynamical friction.  Different models suggest the late
stages are dominated by either gas accretion, or by continued
dynamical friction with stars, but eventually the black holes are
proximate enough that their gravitational attraction takes over.
There is ample evidence for the merger of galaxies through the
observations of binary quasars and x-jet morphology, as well as known
super-massive black hole systems that show periodic variations in
brightness, but no known sources of ultra-low-frequency gravitational
waves.  One likely candidate is the BL Lac object, OJ 287, considered
as a baseline source for the calculations in this work.  Future
long-term synoptic surveys, such as those envisioned for the ongoing
observation campaigns of the LSST \cite{2008arXiv0805.2366I}, as well
as efforts to recover historical photometric data such as DASCH
\cite{2012IAUS..285...29G}, could open the pathway to robust and
ongoing mutli-messenger astronomy with PTAs.

\subsection{Gravitational wave detectors}\label{sub.detectors}
In order to understand the capabilities for using propagation delays
between purportedly null signals for constraining alternative
theories, it is useful to work with a generalized framework to
describe the frequency dependent sensitivity of gravitational
wave detectors.

The utility of modern gravitational wave detection techniques, like
interferometry and pulsar timing, is that they are \textit{broadband},
with typical sensitivities covering many decades in frequency across
the gravitational wave spectrum.  From the perspective of considering
tests of gravitational theories, broadband detection of gravitational
waves provides two distinct opportunities.  First, many effects can be
elucidated as a source sweeps in frequency.  For the kinds of tests
being considered in this paper, this would include dispersion effects,
where different frequencies of waves propagate with different
wavespeeds.  Second, broadband detection affords a larger selection of
sources that can be observed to test alternative theories.
Astrophysically, the distribution of the physical parameters that
characterize a source spans wide ranges in values, which in turn means
that sources imprint in the gravitational wave spectrum at different
frequencies; broadband detectors can detect and characterize a wide
range of astrophysically relevant parameters.

Generically, broadband gravitational wave detection methods have a
``bucket'' shaped sensitivity.  The overall shape and level of the
sensitivity bucket is given by the sources of noise that limit the
performance of a detection technique, and by a frequency-dependent
response function that characterizes how frequencies and amplitudes
that are incident on the detector and mixed and transformed into the
output of the detector.  The development of a correct, parameterized
description of the sensitivity curve for a given detector has been
worked out \textit{in extenso} for each of the detection schemes
considered in this paper, including pulsar timing
\cite{2011PhRvD..83h1301J}, space-based observatories that follow the
basic LISA architecture \cite{2002PhRvD..66f2001L}, as well as
ground-based detectors like LIGO \cite{2012arXiv1203.2674T}.

Sensitivity curves of the sort usually depicted in the literature are
not suited to fully-realized data analysis; they are typically
averaged over the entire sky and over source orientations to provide
an ``average sensitivity'' that is good for developing broad
strategies for utilizing gravitational wave data to answer specific
scientific questions.  For the work presented here, fits to the
standard sensitivity curve buckets are used, rather than curves
completely parameterized for all the variables that can affect
performance of a given detection technique.  The fits presented here
are given as a function of frequency, and utilized to express limits
on alternative theories in a general way that will be applicable to
many astrophysical scenarios.  Sensitivity curves are generally
computed from $S_{N}(f)$, the ``strain spectral density'' as a
function of gravitational wave frequency $f$; the bucket shaped
sensitivity curve that is usually depicted is the square root of the
power spectral density, $h_{f}(f) = \sqrt{S_{N}(f)}$.

The overall shape of the bucket is a function of the noise in a
detector, and its frequency dependent response function to
gravitational waves.  In the high frequency band of the gravitational
wave spectrum ($10 \mbox{Hz} \lesssim f \lesssim 1000 \mbox{Hz}$,
observed by LIGO and similar ground-based detectors), laser
interferometric detectors are limited at low-frequencies by seismic
and gravitational gradient noise, at intermediate frequencies by
thermal noise and noise associated with optical design and coatings,
and at higher frequencies by quantum noise.  In the milli-hertz band
($10^{-5} \mbox{Hz} \lesssim f \lesssim 0.01 \mbox{Hz}$, covered by
space-based LISA-like detectors), the low frequency is dominated by
acceleration noise associated with impulses from the
space-environment, at mid-frequencies by position measurement noises
(dominated by the laser-shot noise), and at high frequencies
sensitivity falls off when the arms of the interferometer are longer
than a gravitational wavelength.  In the nano-hertz band, ($10^{-9}
\mbox{Hz} \lesssim f \lesssim 10^{-6} \mbox{Hz}$, covered by pulsar 
timing arrays), has red ``timing noise'' at low frequencies (noise 
associated with variations in the measured arrival time of the pulsar 
pulses at the detector), white noise associated with the receiver, 
and a loss of sensitivity at high frequencies as the gravitational 
wave periods become shorter than the total span of measured pulsar 
residuals in time. The hard-cutoff at low frequencies is at 
frequencies with periods longer than the total span of data covered 
by pulsar timing.

In this work, fits to the power spectral density, $S_{N}(f)$, were
used to provide simple, frequency dependent fucntions that could be
easily manipulated in an algebraic fashion \cite{2009LRR....12....2S}.
For each detection technique considered, the noise PSD is given in
terms of a dimensionless frequency $x = f/f_{0}$ and rises steeply
above a lower cutoff $f_{s}$.

For the Advanced LIGO noise curve, parameters are chosen to be
$f_{s} = 20$ Hz, $f_{0} = 215$ Hz, and $S_{0} = 1.0 \times 10^{-49}$ 
Hz$^{-1}$, with a fitting formula of
\begin{equation}
    S_{N}(x)/S_{0} = x^{-4.14} - 5x^{-2} + \frac{ 111 (1 - x^2 + 
    0.5 x^4)} {1 + 0.5 x^2}
    \label{eqn.aligoFit}
\end{equation}

For the classic LISA sensitivity curve, parameters are chosen to be
$f_{s} = 10^{-5}$ Hz, $f_{0} = 10^{-3}$ Hz, and $S_{0} = 9.2 \times
10^{-44}$ Hz$^{-1}$, with a fitting formula of
\begin{equation}
    S_{N}(x)/S_{0} = (x/10)^{-4} + 173 + x^{2}
    \label{eqn.lisaFit}
\end{equation}

For PTA sources we use the SKA sensitivity curve from \footnote{http://rhcole.com/apps/GWplotter}. Details can be found in \cite{Moore:2014lga}. The parameters are chosen to be
$f_{s} = 10^{-9}$ Hz, $f_{0} = 10^{-8}$ Hz, and $S_{0} = 1 \times
10^{-24}$ Hz$^{-1}$, with a fitting formula of
\begin{equation}
    S_{N}(x)/S_{0} = 7.04x+(1.81\times10^{-4})x^{-4.5}.
    \label{eqn.ptaFit}
\end{equation}


\subsection{Comparing light and gravitational waves}\label{sub.emGW}
Throughout this work the assumption will be made that $v_{em} = c$.
Is this assumption justified? This consideration has been outlined 
before \cite{Larson:2000p2937}.  If $v_{em} < c$, then photon
propagation would be delayed by some mechanism, making it all but
impossible to construct a meaningful comparison of the two propagation
speeds suggested by Eq.\ \ref{eqn.delta}.  Consider the possibility of
slow photons in the context of the photon having a putative mass.  The
relativistic energy of a massive particle is $E^{2} = p^{2}c^{2} +
m^{2}c^{4}$, so the velocity of the particle will be
\begin{equation}
	\left(\frac{v}{c}\right)^{2} = 1 - \frac{m^{2}c^{4}}{E^{2}}\ .
	\label{eqn.massiveSpeed}
\end{equation}
If we assume the particles travel very close to $c$, so $m \ll E$, we
can write
\begin{equation}
	\varepsilon = 1 - \frac{v}{c} \simeq
	\frac{1}{2}\frac{m^{2}c^{4}}{E^{2}}\ .
	\label{eqn.massiveEpsilon}
\end{equation}
Consider existing bounds on the mass of both the photon and the
graviton.  The current bound on the mass of the photon from
propagation experiments is $m_{em} < 2.3 \times 10^{-33}$ eV/c$^{2}$
\cite{2012PhRvD..86a0001B}.  For optical photons, $\lambda = 500$ nm
and $E \sim 2.5$ eV, implying
\begin{equation}
	\varepsilon_{em} \lesssim 3 \times 10^{-33}\ .
	\label{eqn.emEpsilon}
\end{equation}
One can also consider graviton mass bounds in the context of
propagation delays.  For existing solar system bounds on the graviton
mass, $m_{g} < 4.9 \times 10^{-39}$ eV/c$^{2}$.  In the high
frequency band covered by ground-based detectors ($f \simeq 1$ kHz),
then $E_{gw} = hf \simeq 4.1 \times 10^{-12}$ eV, yielding
\begin{equation}
	\varepsilon_{gw} \lesssim 5.8 \times 10^{-21}\ .
	\label{eqn.ligoEpsilon}
\end{equation}
Similarly, for gravitational waves in the low-frequency band covered
by space-based detectors ($f \simeq 1$ mHz) have $E_{gw} = hf = 4
\times 10^{-18}$ eV, and
\begin{equation}
	\varepsilon_{gw} \lesssim 6.1 \times 10^{-9}\ .
	\label{eqn.lisaEpsilon}
\end{equation}

The result of this consideration is that any putative propagation
delay in the electromagnetic signal is currently constrained far
better than similar bounds on propagation delays in gravitational
signals.  Eq.~\ref{eqn.delta} can be recast in terms of $\varepsilon$
as
\begin{equation}
	\delta \geq \frac{|v_{gw} - v_{em}|}{c} = |\varepsilon_{gw} -
	\varepsilon_{em}|\ .
	\label{eqn.deltaEpsilon}
\end{equation}
This suggests that current bounds on propagation effects already
provide good bounds on $\delta$; the remainder of this work explores
how multi-messenger detections of gravitational wave sources will
further constrain $\delta$.

\subsection{Multi-messenger phase comparison}\label{sub.emGWphase}

By simultaneously monitoring a source in the electromagnetic spectrum
and the gravitational wave spectrum, the phase of the gravitational
wave signal may be compared to the phase of the electromagnetic light
curve (either orbital modulation in binaries, or pulse streams for an
isolated pulsar).  The phases of the electromagnetic and gravitational
wave signals will be offset when they arrive at a distant observer's
location in a manner which depends on the specific astrophysical
details of the emitting system.  For circularized binaries, the
gravitational wave signal emission peaks along the axis connecting the
two constituents of the binary; for asymmetric pulsars the 
gravitational wave signal emission peaks when the maximum distortion 
is in a plane along the line of sight.

Generically, for a wave of frequency $f_{i}$ propagating with 
speed $v_{i}$ over a distance $D$ from the source to the observer, 
the \textit{observed phase} is given by
\begin{equation}
    \phi_{i} = 2\pi D f_{i}/v_{i} \ .
    \label{eqn.phaseObserved}
\end{equation}
The phase difference between the EM light curve and the GW waveform 
at any epoch is given by
\begin{equation}
    \Phi = \phi_{gw} - \phi_{em} = 2\pi D \left(\frac{f_{gw}}{v_{gw}} 
    - \frac{f_{em}}{v_{em}}\right) + \alpha
    \label{eqn.phaseDifference}
\end{equation}
where $\alpha$ is a phase offset at the time of emission due to 
differences in the emission process between gravitational waves and 
the electromagnetic light curve.  In general, $\alpha$ is unknown, 
though in some cases (such as eclipsing binaries) it's value can be 
determined from additional constraints on the geometry of the system.

In principle, $\alpha$ has two components --- one due to delays in the
emission model (geometric factors that influence the location of the
electromagnetic emission) and one which is due to propagation delays
experienced by the signals on their flight toward Earth.  Propagation
delays can be characterized by the photons traveling through media
with non-unit index of refraction: first through the interstellar
medium, and then through the Earth's atmosphere.  Given the typical
distance to sources that will be simultaneously detectable in both EM
and GW spectra, the typical phase delay from propagation delays is
estimated to be $\alpha_{prop} \sim 5 \times 10^{-11}$, about 4 orders
of magnitude less than the expected accuracy of the raw phase
measurements themselves \cite{Larson:2000p2937}, and can be safely
neglected for this analysis.

The emission model contribution to $\alpha$ parameterize the initial
phase lag between the gravitational wave and electromagnetic signal.
It is an astrophysical quantity which represents the relative phase
difference between the peaks in the light curve and the peaks in the
quadrupolar gravitational radiation pattern.  In a binary, determining
the value of $\alpha$ requires knowledge of the position of the stars
at the time the light curve peaks.  In a pulsar $\alpha$ will
represent the difference between the beam emission and the
ellipticity that gives the quadrupole moment.

The ability to determine $\alpha$ will be limited by understanding 
the model for electromagnetic emission, and ultimately by the errors
associated with each of the independent phase measurements.

\begin{figure}[t!]
  \centering
 \includegraphics[width=80mm]{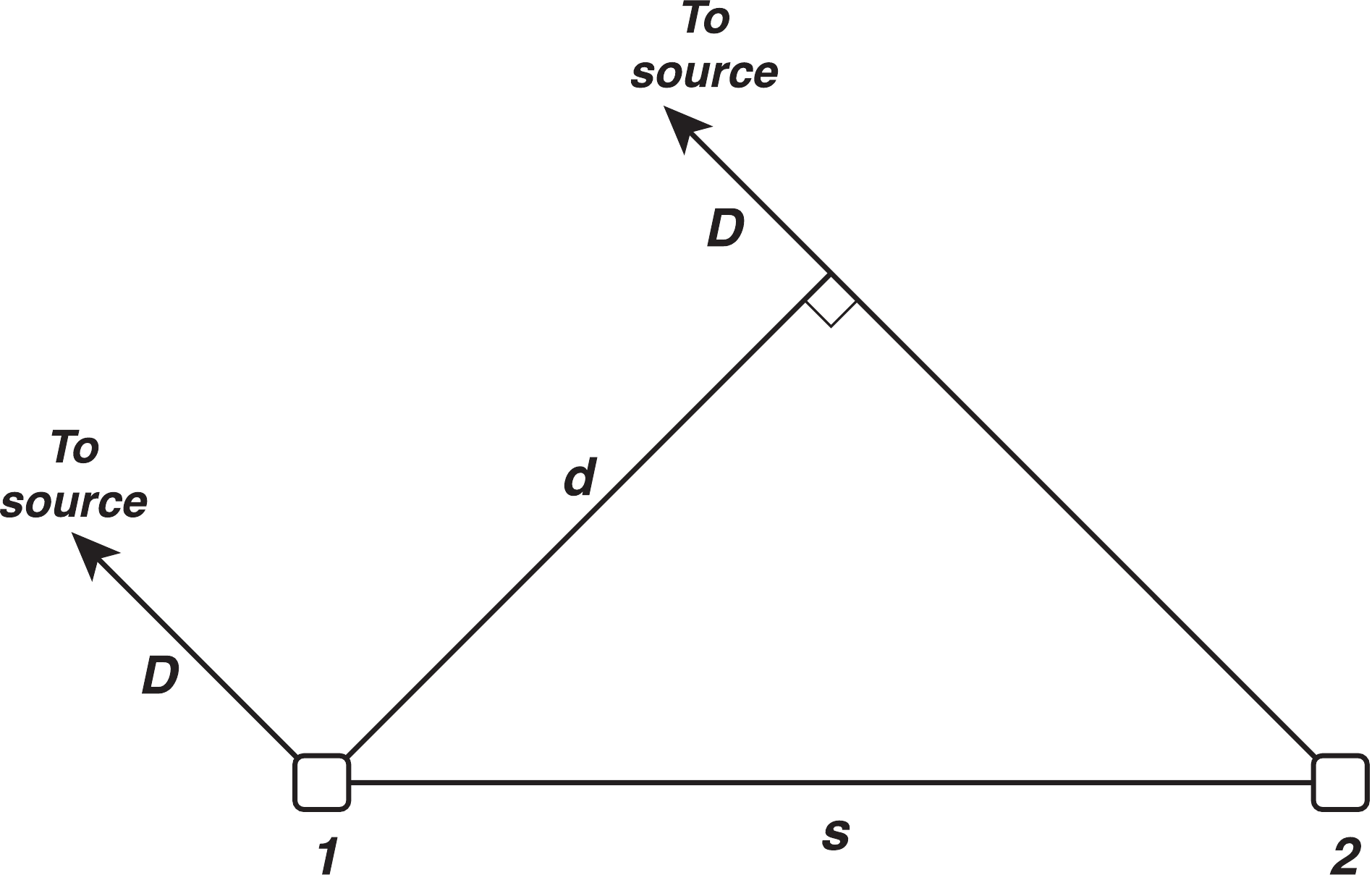} 
  \caption{Observations of a source at two different epochs (or 
  locations), $1$ and $2$. Comparing phase measurements at the two 
  epochs, separated in time by $\tau$, amounts to propagation 
  differences built up over the distance $d$.}
  \label{fig.epochObserving}
\end{figure}

Since $\alpha$ is a constant additional factor in the phase, it can be
eliminated by considering the phases at two different observing
epochs.  Consider Figure \ref{fig.epochObserving}, showing a detector
at two different observing epochs, physically separated by a distance
$s$.  For a network of detectors, the two positions could also be the
location of different detectors observing at the same epoch.  The line
of sight to the source has different path length from each of the two
positions, with the waves propagating a distance $d$ farther to one of
the positions.  The value of $d$ is bounded by the equal path length
case ($d = 0$) and the parallel path length case ($d = s$).  If
observing campaigns from the two positions are separated in time by
$\tau$ (where $\tau = 0$ for two, simultaneously observing detectors),
then the phase difference between the two epochs is
\begin{equation}
    \Phi_{2} - \Phi_{1} = 2\pi d \left(\frac{f_{gw}}{v_{gw}} 
    - \frac{f_{em}}{v_{em}}\right) + 2\pi \tau (f_{gw} - f_{em})\ .
    \label{eqn.epochObserving}
\end{equation}
For all the sources considered in this work (isolated pulsars, and
ultra-compact binaries) the gravitational wave frequency and
electromagnetic frequency are related by $f_{gw} = 2 f_{em}$; in the
case of pulsars, $f_{em}$ is the spin frequency, and in the case of
binaries, $f_{em}$ is the orbital frequency.  Making this constraint 
then yields
\begin{equation}
    \Phi_{2} - \Phi_{1} = 2\pi d f_{em}\left(\frac{2}{v_{gw}} 
    - \frac{1}{v_{em}}\right) + 2\pi f_{em}\tau\ .
    \label{eqn.epochObserving2}
\end{equation}

\subsection{Null experiments}\label{sub.nullExperiments}

To date, there is a strong bias toward the belief that gravitational
radiation is correctly described by general relativity, with
gravitational waves propagating with speed $v_{gw} = v_{em} = c$ on
the background spacetime metric.  As such, measurements looking for
deviations from the predictions of general relativity are expected
only to produce bounds constraining alternative theories.  The
approach taken here will be to interpret model measurements as a
\textit{null experiment} --- consider the case where no deviation will
be detected, and $\delta$ can only be bound by the size of the
experimental errors, within which any putative difference between 
general relativity and an alternative theory can hide.

In this approach, all the errors in the measurement are ascribed to
putative differences between the propagation speeds, $v_{gw}$ and
$v_{em}$.  The origin of the errors, whether they be gravitational
wave phase errors $\delta \phi_{gw}$ or electromagnetic phase errors
$\delta \phi_{em}$, is not critical; limiting bounds on alternative
theories are derived from the total phase error $\delta \phi$ defined
by adding the errors in the individual phases in quadrature:
\begin{equation}
    \delta \phi = \sqrt{\delta \phi_{gw}^{2} + \delta \phi_{em}^{2}}\ 
    .
    \label{eqn.totalPhaseError}
\end{equation}
As shown in \cite{Larson:2000p2937, Cutler:2003p2938} these errors can be used to set a limit on the Compton wavelength of the graviton
\begin{equation}
\lambda_{g}=	\frac{1}{f_{gw}}\sqrt{\frac{1}{2}\left(1+\frac{\pi f_{gw}D}{\delta \phi}\right)}
\end{equation}
 and therefore on the mass of the graviton (see below).
 
Errors in the individual phase measurements are estimated from the 
details of the observing campaigns that will recover the phases.  In 
the case of gravitational wave measurements, the errors are 
proportional to the SNR of the signal over the duration of the 
observation \cite{Cutler:1997ta,Cutler:2003p2938}:
\begin{equation}
    \delta \phi_{gw} = \frac{\beta}{2}\left ( \frac{S}{N}\right )^{-1}.
    \label{eqn.GWphaseError}
\end{equation}
The EM phase error will depend on the errors inherent in 
measuring the time of arrival of the data points that comprise the 
light curve.  Generically, for a timing uncertainty $\delta t$ over a 
light curve period $\tau_{c}$, the EM phase uncertainty will be
\begin{equation}
    \delta \phi_{em} = 2\pi\delta t/\tau_{c}\ .
    \label{eqn.EMphaseError}
\end{equation}
For typical timing errors and periods of interacting binaries in the
millihertz band (like AM CVn), $\delta t \sim 10^{-4}$s, with light
curve periods of $\tau_{c}\sim 1000$ s \cite{1998ApJ...493L.105H},
yielding $\delta \phi_{em} \sim 6 \times 10^{-7}$.  For pulsars, typical
precision in time of arrival measurements are $\delta t \sim 10^{-13}$
s, with pulse periods of $\tau_{c} \sim 0.01$ s, yielding $\delta \phi_{em}
\sim 10^{-11}$
\footnote{http://www.atnf.csiro.au/research/pulsar/psrcat}
\cite{2005AJ....129.1993M}. In general these errors will be much smaller than those from the gravitational sector and can be safely left out of calculations.
  
Using the Compton wavelength as a proxy for the mass, $m=\frac{h}{\lambda c}$, gives 
\begin{equation}
m_{g}=\frac{h}{c}\frac{\sqrt{\beta}}{\pi}\sqrt{\frac{2\pi f_{gw}}{D}}\left(\sqrt{\frac{S}{N}}\right)^{-1} \label{eqn.GenMassGrav}
\end{equation}

We can then write the rms SNR \cite{Cutler:2003p2938,Finn:2000sy} as 
\begin{equation}
    \frac{S}{N} = h_{0} \left(2T \right)^{\frac{1}{2}}/\left[S^{SA}_{h}(f_{gw})\right]^{\frac{1}{2}}
    \label{eqn.SNR_SpectralNoise}
\end{equation}
where $S^{SA}_{h}(f_{gw})$ is the spectral noise density of a given detector as a function of frequency. Combining
Eqs. \ref{eqn.GenMassGrav} and \ref{eqn.SNR_SpectralNoise} we get 
\begin{equation}
m_{g}=\frac{h}{c}\frac{\sqrt{\beta}}{\pi}\sqrt{\frac{2\pi f_{gw}}{D}}\frac{1}{\sqrt{h_{0}}}\left(\frac{S^{SA}_{h}(f_{gw})}{2T}\right)^{\frac{1}{4}} \label{eqn.GenMassGravFull}
\end{equation}

\section{Pulsars as multi-messenger sources}\label{sec.gwPulsar}
As sensitivity improves, ground-based gravitational wave detectors
like LIGO and Virgo have been placing stricter limits on the
ellipticity of known pulsars using null detection (upper bounds on
ellipticity fixed by not detecting the pulsars in the gravitational
wave data stream) \cite{Abbott:2007ce,ligo116pulsarsS5,Collaboration:2008p2939}.  The known data from
electromagnetic observations of these pulsars includes the rotational
frequency, mass, and inclination, all of which are needed to estimate
the amplitude of any putative gravitational wave signal.  However we
cannot directly measure the principle moment of inertia, and therefore
cannot fully predict the strain.  For the purposes of this work, the
critical information is the phase of the gravitational waves, which
would be compared to the phase of an electromagnetic signal;
constraints on a gravitational wave phase measurement can be estimated
from the known data.  The presumption in this approach is that the
result will also be a null detection, and the accumulated error from
the measurement process can be attributed to the propagation effect
under consideration, a difference in wave speed between the
gravitational waves and the electromagnetic waves emitted by the
pulsar.

For demonstration here, one can model the eccentricity of the pulsar
as a bump on its spin equator, assuming that the spin axis is a
principal axis.  For the two polarizations states, this gives
amplitudes \cite{Maggiore:1900zz}
\begin{eqnarray}
   h_{+}&=& h_{0}\frac{1+\cos^{2}\iota}{2}\cos\left(2\pi
   f_{gw}t\right) \\
   h_{\times}&=&h_{0}\cos\iota\sin\left(2\pi f_{gw}t\right)\ ,
   \label{eqn.pulsarAmplitudes}
\end{eqnarray}
where $h_{0}$ is a scaling amplitude written in terms of the pulsar
ellipticity $\epsilon$ as
\begin{equation}
  h_{0}=\frac{4\pi^{2}G}{c^{4}}\frac{I_{3}f_{gw}^{2}}{D}\epsilon\ .
  \label{eqn.scalingAmplitudePulsar}
\end{equation}
Here we do not include the spin-down of these pulsars since the error
introduced over the three year observation period by the change in
frequency is at most 3 orders of magnitude smaller than the
gravitational error itself.


Using LIGO over a three year measurement period and using the error in
gravitational wave phase to calculate the maximum value of $\delta$ we
get
\begin{eqnarray}
   \delta &=&\beta\;1.105\times10^{22} \left (S_{h}\times1Hz
   \right)^{\frac{1}{2}} \!  \left (\frac{f_{gw}}{1 Hz} \right)^{\!3}
   \nonumber \\
   &\times & \!\!  \left (\frac{I_{3}}{10^{38}kg\;m^{2}}
   \right)^{-1} \!\!\!  \left (\frac{e}{10^{-5}}
   \right)^{-1} \!\!  \left (\frac{T}{1yr}
   \right)^{-\frac{1}{2}}\ .
   \label{eqn.PulsarDelta}
\end{eqnarray}
where the factor of $\beta$ is a multiplicative factor that includes all the same information as the parameter $\alpha$, $\beta\equiv1+\alpha$ \footnote{It should be pointed out that the $\alpha$ in \cite{Larson:2000p2937} and this work are the same. The $\alpha$ in \cite{Cutler:2003p2938} is what we refer to here as $\beta$.}. Assuming instead that the discrepancy from the speed of light reveals
a graviton with non-zero mass the error shows that the graviton mass,
calculated from the LIGO error curve is
\begin{eqnarray}
   m_{g} &=& \sqrt{\beta}\;3.95\times10^{-25} \left (S_{h}\times1Hz
   \right)^{\frac{1}{4}} \left (\frac{f_{gw}}{1 Hz}
   \right)^{-\frac{1}{2}} \nonumber \\
   &\times& \!\!\left (\!\frac{I_{3}}{10^{38}kg\;m^{2}}
   \!\right)^{-\frac{1}{2}} \!\!\!\left
   (\!\frac{e}{10^{-5}}\!\!  \right)^{-\frac{1}{2}} \!\!  \left
   (\frac{T}{1yr} \right)^{-\frac{1}{4}} \!\!\frac{eV}{c^{2}}\ 
   .
   \label{eqn.PulsarMassLimit}
\end{eqnarray}

When taken as a function of frequency, there is a minimum in the
graviton mass, while $\delta$ continues to decrease
as the frequency of the gravitational waves increases.  The
calculation can of course be done for any pulsar where the frequency
is known, but we limit our scope to the 116~pulsars included in
\cite{ligo116pulsarsS5}.  We have calculated the limits that
would be given by the 116 pulsars from \cite{ligo116pulsarsS5} in
Appendix 1.

\section{Binaries as multi-messenger sources} \label{sec.gwBinary}
The ultra-compact binaries are one of the principal sources that will
be detected by space-based gravitational wave interferometers.  It is
expected that a LISA class mission will detect on the order of
$10^{4}$ individually resolved ultra-compact binaries
\cite{2007CQGra..24S.575C}, of which $\sim 10^{2}$ or more will be
simultaneously observable in gravitational waves and with
electromagnetic telescopes \cite{2013MNRAS.429.2361L}.  Foremost among
these will be a group of ultra-compact binaries that have already been
detected electromagnetically, and are expected to be strong sources of
gravitational wave emissions that will be detected very soon after a
space-based observatory becomes operational; these are collectively
known as \textit{verification binaries} \cite{NelemansWiki}.  The
frequency band of interest is populated by systems that have orbital
periods on the order of several thousand seconds to tens of seconds,
and the gravitational wave emission is well described by the
quadrupole formula \cite{PM,PM63}.  The gravitational wave emission
extracts energy and angular momentum from the binary on long
timescales, until ultimately the components merge (for compact stellar
remnants like neutron stars and black holes, the merger occurs at high
frequencies, in the regime covered by ground-based gravitational wave
detectors).  The overall strength of the gravitational waves depends
on a scaling factor $h_{o}$:
\begin{equation}
	h_{o} = \frac{4G^{2}m_{1}m_{2}}{c^{4}a(1 - e^{2})D} =
	\frac{4G^{5/3}}{c^{4}(1-e^{2})}\frac{{\cal
	M}}{D}\left(\frac{1}{2} \pi f_{o} {\cal M}\right)^{2/3}
    \label{eqn.scalingAmplitude}
\end{equation}
where $f_{o}$ is the orbital frequency of the binary, $D$ is the
luminosity distance, and ${\cal M} = (m_{1}m_{2})^{3/5}/(m_{1} +
m_{2})^{1/5}$ is the ``chirp mass'' of the system.  Using a
space-based interferometer over a three year measurement period and
using the error in gravitational wave phase from
Eq.\ref{eqn.SNR_SpectralNoise} to calculate the maximum value of
$\delta$ we get

\begin{eqnarray}
   \delta &=&\beta\;8.77\times10^{-13}
   \frac{\left(M_{1}+M_{2}\right)^{\frac{1}{3}}}{M_{1}M_{2}}\left(\frac{10^{-3}Hz}{f_{0}}\right)^{\frac{11}{6}}\nonumber
   \\
   &&\times \left(\frac{T}{1\;
   yr}\right)^{-\frac{1}{2}}\left(S_{h}^{SA}\times1Hz\right)^{\frac{1}{2}}
   \label{eqn.BinaryDelta}
\end{eqnarray}
If instead we choose to characterize the discrepancy as a massive graviton we have

\begin{eqnarray}
	m_{g}
	&=&\sqrt{\beta}\;1.22\times10^{-40}\frac{\left(M_{1}+M_{2}\right)^{\frac{1}{6}}}{\sqrt{M_{1}M_{2}}}
	\left(\frac{10^{-3}Hz}{f_{0}}\right)^{-\frac{1}{6}} \nonumber \\
	&& \times \left(\frac{T}{1\; yr}\right)^{-\frac{1}{4}}\left(S_{h}^{SA}\times1Hz\right)^{\frac{1}{4}}\frac{eV}{c^{2}}
	\label{eqn.PulsarMassLimit}
\end{eqnarray}

For the analysis here, it is assumed the binaries are
\textit{monochromatic} over the duration of the measurement.
Generically, the contribution of a frequency derivative $\dot f$ will 
only be important for binaries that evolve in frequency by an amount 
${\dot f}\cdot \Delta T \simeq \delta f = 1/T_{obs}$, where $\delta 
f$ is the frequency resolution of the observation.

\section{Pulsar timing arrays and multi-messenger astronomy}\label{sec.gwPTA}
Multi-messenger astronomy with pulsar timing arrays (PTAs) is a 
largely unexplored avenue of observation, but in concept is 
identical to multi-messenger observing campaigns described in 
Sections \ref{sec.gwPulsar} and \ref{sec.gwBinary} for higher 
frequency bands; the timescales with a PTA are simply longer.

Just as with the ultra-compact binaries, PTA multi-messenger 
astronomy requires a reconstructed phase curve for both the 
electromagnetic and gravitational wave observations. Most monitoring 
programs are carried out on a regular schedule (for both EM 
observations and pulsar timing observations), but at irregularly 
spaced time intervals, and are not necessarily coincident with pulsar 
timing observations. However, over long periods of time, repeated 
sampling will map out the signals, enabling a multi-messenger phase 
comparison.

The synergy of multi-messenger astronomy with PTAs has been 
demonstrated before when a null-detection of gravitational waves was 
used to rule out a binary explanation for reported radio intensity 
oscillations of the quasar 3C66b \cite{2004ApJ...606..799J}. There 
are certainly known binary super-massive black holes (SMBHs) that 
have observed EM light curves and orbital periods firmly in the PTA 
observing band, though current sensitivities are not sufficient to 
detect their gravitational wave emission. One such example (used as a 
reference point in this paper) is quasar OJ 287, which has a 100-year 
baseline of electromagnetic observations \cite{2011ChAA..35..123S}.

The prospects for using multi-messenger astronomy with PTAs will be 
constrained by the overall population of SMBH binaries, as it is 
expected their signals will be overlapping and form a stochastic 
background in the pulsar timing band, similar to the expected 
confusion foreground from galactic binaries in the millihertz band 
observed by LISA-like detectors. Early work based on observed galaxy 
merger rates have produced useful general formulae for the expected 
number of merging binaries as a function of redshift and cosmological 
model, which can be used to estimate the prospective number of 
sources in the PTA band \cite{2003ApJ...583..616J, 
2000ApJ...532L...1C, 2002ApJ...565..208P}.  More recent work has 
sought to build different estimates based on numerical simulations of 
merger trees, built from theoretical considerations  (e.g.
\cite{2001ApJ...558..535M, 2004ApJ...611..623S, 
2007MNRAS.380.1533M,2013MNRAS.433L...1S}), though these models do not 
have the same useful algebraic form as the phenomenological models in 
\cite{2003ApJ...583..616J}.

\section{Discussion}\label{sec:Discussion}
Once regular gravitational wave observations commence, and 
gravitational wave data becomes openly available to the astronomical 
community, multi-messenger observing campaigns for known, targeted 
sources will become possible.  Similarly, as electromagnetic 
surveys continue to expand new and interesting multi-messenger 
targets will be discovered that will expand the suite of targets that 
can be used to bound alternative theories using the propagation 
analysis presented here.  

\begin{figure}[h]
\begin{center}
\includegraphics[width=4.5in]{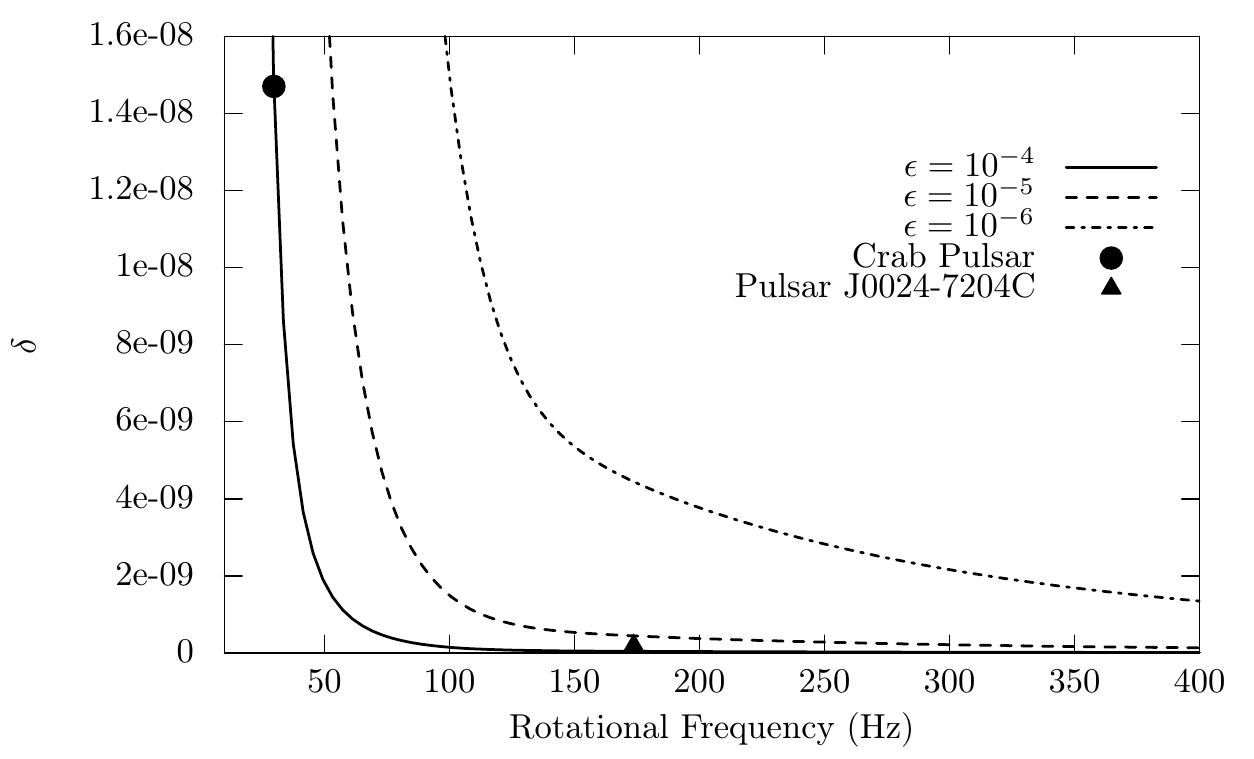}
\includegraphics[width=4.5in]{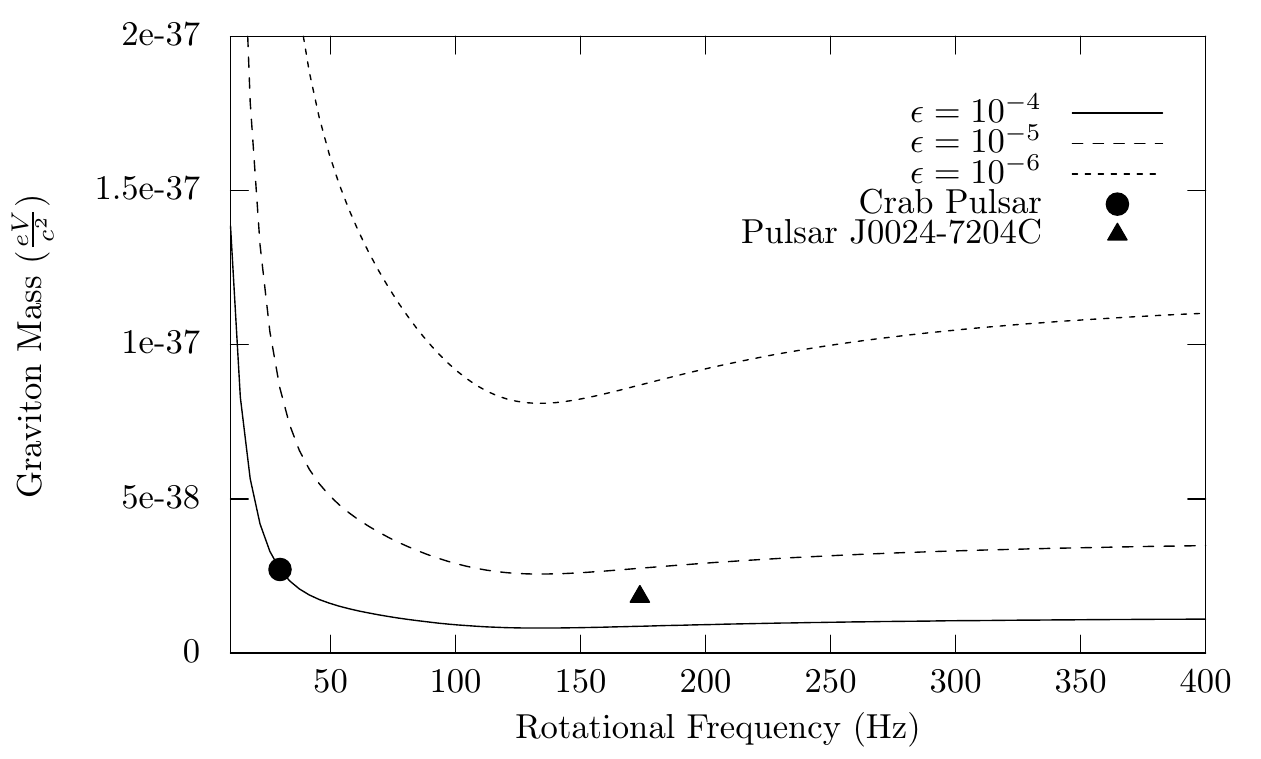}

\caption{The top panel shows the maximum difference from the speed of
light, $\delta$, and the lower panel shows the maximum graviton mass,
$m_{g}$ that can be attributed to the error in phase measurement from
a pulsar, given three different values of the ellipticity.  The limits
possible from the Crab pulsar and J0024-7204C are plotted given the
ellipticity limits from \cite{ligo116pulsarsS5}.}
\label{fig.PulsarPlots}

\end{center}
\end{figure}

Given the possible ranges of astrophysical parameters that can 
reasonably expected to be covered by future discoveries, the analysis 
here can by summarized by bounding curves on the maximum deviation in 
propagation speed from the speed of light, $\delta$ (Eq.\ 
\ref{eqn.delta}), and the graviton mass $m_{g}$.  

\begin{figure}[h]
\begin{center}
\includegraphics[width=4.5in]{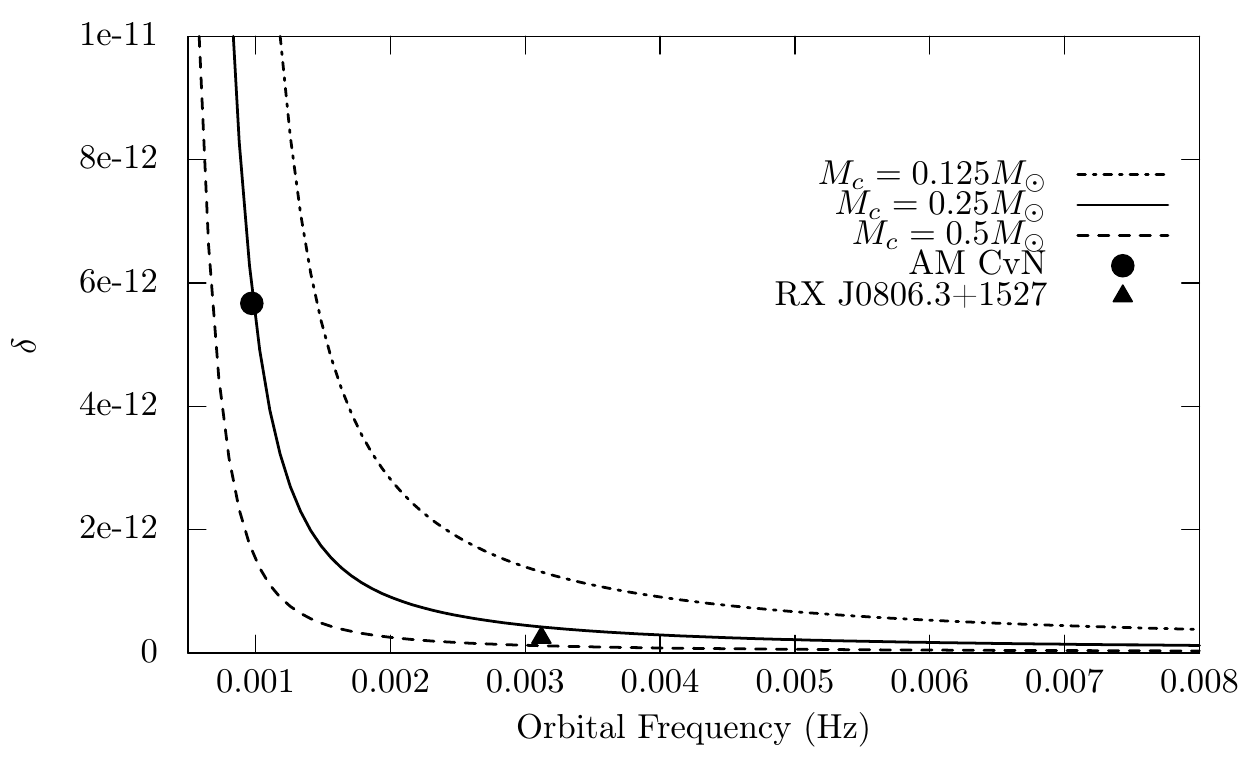}
\includegraphics[width=4.5in]{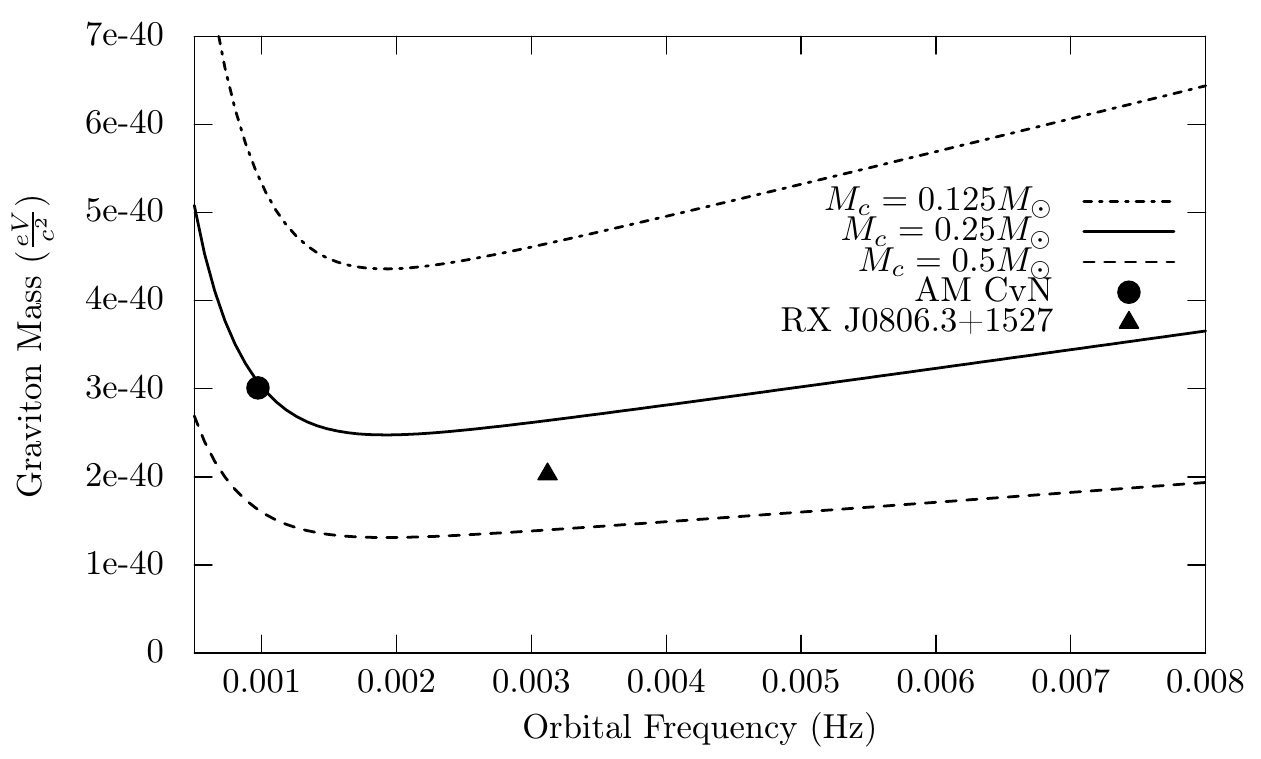}

\caption{The top panel shows the maximum difference from the speed of
light, $\delta$, and the lower panel shows the maximum graviton mass,
$m_{g}$ that can be attributed to the error in phase measurement from
a binary system, given three different values of the chirp mass.}
\label{fig.BinaryPlots}

\end{center}
\end{figure}

In Figure \ref{fig.PulsarPlots} we have drawn the relationship between
the frequency of the pulsar and the target quantities in this
manuscript; the difference from the speed of light in the upper panel and
the mass of the graviton in the lower panel.  The curves have been
drawn for three different values of the ellipticity, against the
standard LIGO spectral density curve for the canonical value of $I_{3}$ ($10^{38}kg\;m^{2}$), fixed observation time ($T=3\;yrs$) and negligible $\beta$.
Generically, one sees that much better bounds on $\delta$ are obtained
from higher spin frequencies.  The dependence on ellipticity is a
manifestation of the dependence of the gravitational wave emission on
the star being deformed; there is stronger emission from strong
deformations, leading to higher SNR, thus as the ellipticity increases
the bound on $\delta$ increases.  Similarly, one can see that the
ellipticity is the more important factor for putting a limit on the
graviton mass at higher frequencies in the band.  The expected limits from 
non-detection of known sources \cite{ligo116pulsarsS5} (the Crab Pulsar and PSR J0024-7024C) 
are indicated.

\begin{figure}[h]
\begin{center}
\includegraphics[width=4.5in]{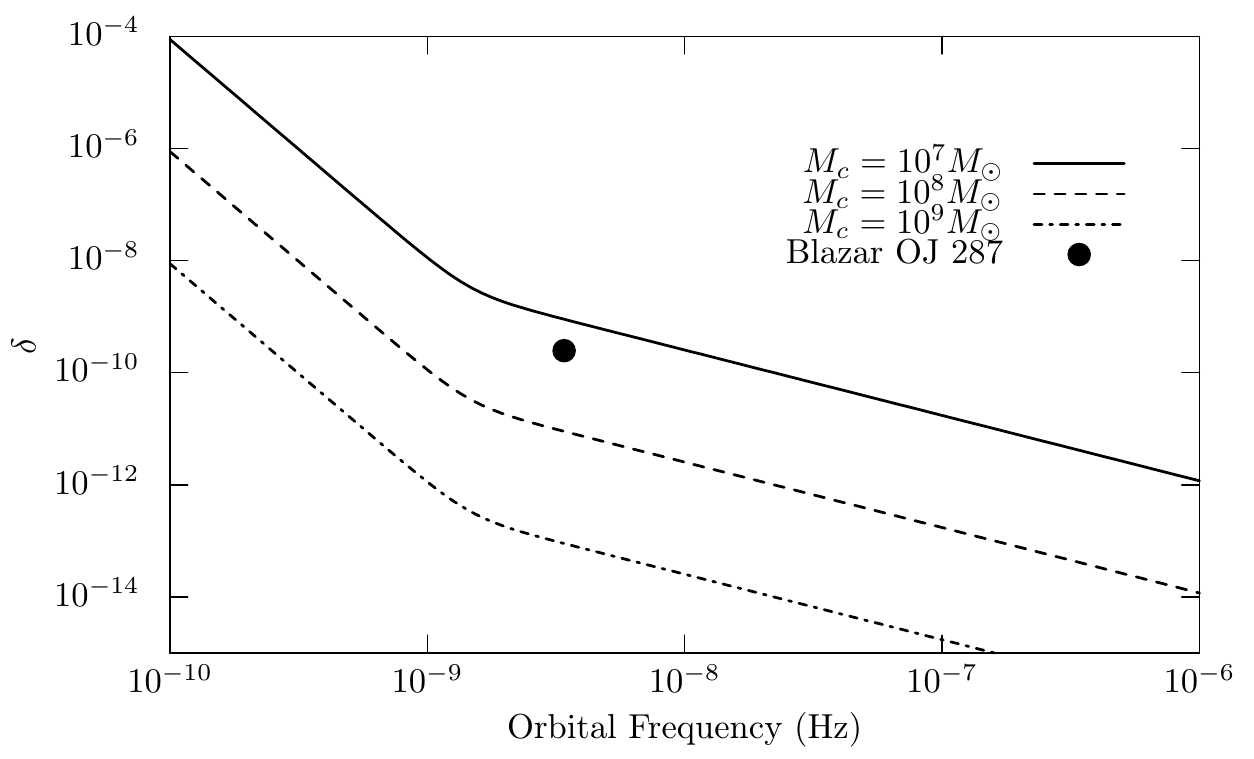}
\includegraphics[width=4.5in]{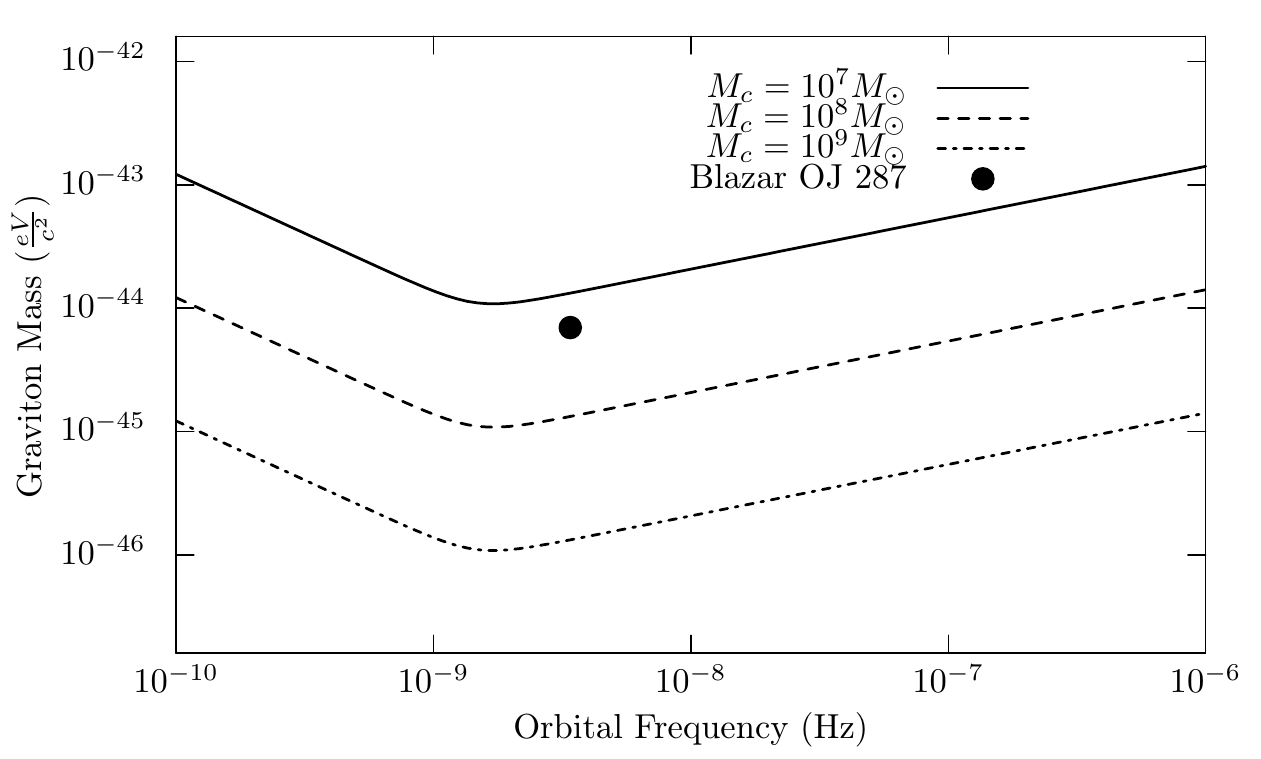}

\caption{The top panel shows the maximum difference from the speed of
light, $\delta$, and the lower panel shows the maximum graviton mass,
$m_{g}$ that can be attributed to the error in phase measurement from
a binary quasar system, given three different values of the chirp mass.}
\label{fig.BinaryQuasarPlots}

\end{center}
\end{figure}

Similarly in Figure \ref{fig.BinaryPlots} we have drawn the
relationship between the orbital frequency of an ultra-compact binary
and the quantities of concern in this manuscript; the difference from
the speed of light in the upper panel and the mass of the graviton in the
lower panel.  The curves have been drawn for three different values of
the chirp mass, ${\cal M}$, again for the standard LISA spectral density
curve and for fixed observation time and negligible $\beta$. As was the case with the pulsars, overall
bounds are improved by increasing the SNR, so higher mass systems
produce more stringent bounds.  The expected bound from the detection
of known systems (AM CVn and RX J0806.3+1527) are indicated.

Figure~\ref{fig.BinaryQuasarPlots} represents the limits that we can put on $m_{g}$ and $\delta$ using binary quasars as a source. Here we have again drawn three curves for three different values of chirp mass, $\mathcal{M}$, here using the sensitivity curve from \cite{Moore:2014lga} again for fixed time (3 years) and negligible $\beta$. Note that the limit on $m_{g}$ from observations involving Blazar OJ 287 would be the most stringent limits of any of the specific sources cited herein.

These curves are representative of the null interpretation used in this
paper; for a source in the space of these plots, the point is an upper
bound on the deviations from general relativity produced by the
alternative theory.  Improved accuracy in phase measurements, or
direct gravitational wave detections of a source that firmly establish
currently unknown parameters (such as the ellipticity of a known
pulsar) will further constrain the alternative theories by pushing 
the upper bounds to lower values.

Lastly, one should note that the work presented here is a proof of
concept, demonstrating the bounds on alternative theories that could
be derived from current best results and reasonable predictions for
our ability to conduct multi-messenger observing campaigns. There are both theoretical and logistical issues that we choose not to consider here. From a theoretic standpoint it should be noted that in at least one bimetric theory \cite{Mohseni:2011p3060, Mohseni:2012p3424} only the dilatational mode (not found in GR and not detectable by single interferometers) moves at a propagation speed slower than c.
The observational logistical details will only emerge after the nascent
multi-messenger partnerships between the electromagnetic and
gravitational wave communities grow into full fledged coordinated
observing programs.

\section*{Acknowledgments} SLL acknowledges support from National Science 
Foundation award PHY-0970152, and from NASA award NNX13AM10G.

\section*{Appendix}\label{App1}
\begin{center}
{\small

\begin{filecontents*}{116GravLimitsTableA.csv}
Pulsar,Frequency (Hz),Ellip LIGO,delta div alpha,graviton mass (eV)
J0024-7204C ,173.71,2.26e-05,4.45e-10,2.47e-21
J0024-7204D ,186.65,1.48e-06,4.09e-10,2.55e-21
J0024-7204E ,282.78,1.44e-06,2.37e-10,2.94e-21
J0024-7204F ,381.16,6.98e-07,1.46e-10,3.12e-21
J0024-7204G ,247.5,1.90e-06,2.87e-10,2.83e-21
J0024-7204H ,311.49,7.69e-07,2.04e-10,3.01e-21
J0024-7204I ,286.94,7.30e-07,2.32e-10,2.95e-21
J0024-7204J ,476.05,5.34e-07,9.91e-11,3.20e-21
J0024-7204L ,230.09,1.27e-06,3.17e-10,2.76e-21
J0024-7204M ,271.99,9.61e-07,2.51e-10,2.91e-21
J0024-7204N ,327.44,9.02e-07,1.89e-10,3.04e-21
J0024-7204Q ,247.94,1.08e-06,2.86e-10,2.83e-21
J0024-7204R ,287.32,7.76e-07,2.31e-10,2.95e-21
J0024-7204S ,353.31,6.33e-07,1.66e-10,3.08e-21
J0024-7204T ,131.78,2.23e-06,6.71e-10,2.31e-21
J0024-7204U ,230.26,1.23e-06,3.16e-10,2.77e-21
J0024-7204Y ,455.24,5.26e-07,1.07e-10,3.19e-21
J0218+4232 ,430.46,1.10e-06,1.19e-10,3.17e-21
J0407+1607 ,38.91,3.93e-05,4.71e-08,5.71e-21
J0437-4715 ,173.69,6.74e-07,4.45e-10,2.47e-21
J0534+220,29.8,1.00e-04,1.47e-07,7.72e-21
J0613-0200 ,326.6,1.18e-07,1.89e-10,3.03e-21
J0621+1002 ,34.66,5.65e-05,7.58e-08,6.44e-21
J0711-6830 ,182.12,3.71e-07,4.21e-10,2.52e-21
J0737-3039A ,44.05,1.10e-05,2.92e-08,5.08e-21
J0751+1807 ,287.46,2.91e-07,2.31e-10,2.95e-21
J1012+5307 ,190.27,2.36e-07,4.00e-10,2.57e-21
J1022+1001 ,60.78,1.14e-06,9.08e-09,3.91e-21
J1024-0719 ,193.72,1.67e-07,3.92e-10,2.59e-21
J1045-4509 ,133.79,1.87e-06,6.50e-10,2.30e-21
J1455-3330 ,125.2,5.75e-07,7.53e-10,2.32e-21
J1600-3053 ,277.94,4.55e-07,2.43e-10,2.93e-21
J1603-7202 ,67.38,1.98e-06,6.28e-09,3.61e-21
J1623-2631 ,90.29,3.71e-06,2.16e-09,2.84e-21
J1640+2224 ,316.12,1.87e-07,1.99e-10,3.01e-21
J1643-1224 ,216.37,1.07e-06,3.43e-10,2.70e-21
J1701-3006A ,190.78,2.61e-06,3.99e-10,2.57e-21
J1701-3006B ,278.25,1.61e-06,2.43e-10,2.93e-21
J1701-3006C ,131.36,3.32e-06,6.75e-10,2.31e-21
J1713+0747 ,218.81,2.45e-07,3.38e-10,2.72e-21
J1730-2304 ,123.11,4.72e-07,7.85e-10,2.33e-21
J1732-5049 ,188.23,6.34e-07,4.05e-10,2.56e-21
J1744-1134 ,245.43,2.07e-07,2.90e-10,2.82e-21
J1748-2446A ,86.48,6.77e-06,2.53e-09,2.94e-21
J1748-2446C ,118.54,4.63e-06,8.69e-10,2.36e-21
J1748-2446D ,212.13,1.96e-06,3.51e-10,2.68e-21
J1748-2446E ,455,5.62e-07,1.07e-10,3.19e-21
J1748-2446F ,180.5,3.34e-06,4.25e-10,2.51e-21
J1748-2446G ,46.14,3.56e-05,2.45e-08,4.88e-21
J1748-2446H ,203.01,2.46e-06,3.70e-10,2.64e-21
J1748-2446I ,104.49,4.21e-06,1.29e-09,2.53e-21
J1748-2446K ,336.74,7.65e-07,1.80e-10,3.05e-21
J1748-2446L ,445.49,9.09e-07,1.12e-10,3.18e-21
J1748-2446M ,280.15,1.68e-06,2.40e-10,2.93e-21
J1748-2446N ,115.38,5.71e-06,9.38e-10,2.39e-21
J1748-2446O ,596.43,9.68e-07,6.54e-11,3.26e-21
J1748-2446P ,578.5,6.08e-07,6.93e-11,3.25e-21
\end{filecontents*}
\pgfplotstabletypeset[
	col sep=comma,
    	string type,
	columns/Pulsar/.style={column name=Pulsar, column type={| l}},
	columns/Frequency (Hz)/.style={column name= $f_{em}\;(Hz)$, column type={| r |}},
	columns/Ellip LIGO/.style={column name=$\epsilon$, column type={c |}},
	columns/delta div alpha/.style={column name= $\delta \Big/\beta$, column type={c |}},
	columns/graviton mass (eV)/.style={column name= $m_{g}\Big/\sqrt{\beta}\;(\frac{eV}{c^{2}})$, column type={c |}},
	every head row/.style={before row=\hline,after row=\hline},
	every last row/.style={after row=\hline},
	]{116GravLimitsTableA.csv}
	
\begin{filecontents*}{116GravLimitsTableB.csv}
Pulsar,Frequency (Hz),Ellip LIGO,delta div alpha,graviton mass (eV)
J1748-2446Q ,355.62,9.05e-07,1.65e-10,3.08e-21
J1748-2446R ,198.86,2.71e-06,3.80e-10,2.62e-21
J1748-2446S ,163.49,2.17e-06,4.78e-10,2.41e-21
J1748-2446T ,141.15,3.34e-06,5.89e-10,2.31e-21
J1748-2446V ,482.51,7.04e-07,9.67e-11,3.20e-21
J1748-2446W ,237.8,2.20e-06,3.03e-10,2.80e-21
J1748-2446X ,333.44,9.57e-07,1.83e-10,3.05e-21
J1748-2446Y ,488.24,1.15e-06,9.46e-11,3.21e-21
J1748-2446Z ,406.08,6.65e-07,1.31e-10,3.14e-21
J1748-2446aa ,172.77,9.92e-06,4.47e-10,2.47e-21
J1748-2446ab ,195.32,1.59e-06,3.88e-10,2.60e-21
J1748-2446ac ,196.58,2.42e-06,3.85e-10,2.60e-21
J1748-2446ad ,716.36,4.48e-07,4.62e-11,3.29e-21
J1748-2446ae ,273.33,1.14e-06,2.49e-10,2.91e-21
J1748-2446af ,302.63,1.52e-06,2.14e-10,2.99e-21
J1748-2446ag ,224.82,2.44e-06,3.26e-10,2.74e-21
J1748-2446ah ,201.4,1.76e-06,3.74e-10,2.63e-21
J1756-2251 ,35.14,5.42e-05,7.15e-08,6.35e-21
J1801-1417 ,275.85,3.44e-07,2.46e-10,2.92e-21
J1803-30 ,140.83,5.12e-06,5.91e-10,2.31e-21
J1804-0735 ,43.29,8.95e-05,3.12e-08,5.16e-21
J1804-2717 ,107.03,5.79e-07,1.19e-09,2.49e-21
J1807-2459A ,326.86,9.13e-07,1.89e-10,3.04e-21
J1810-2005 ,30.47,2.28e-04,1.33e-07,7.51e-21
J1823-3021A ,183.82,2.17e-06,4.16e-10,2.53e-21
J1824-2452A ,327.41,8.43e-07,1.89e-10,3.04e-21
J1824-2452B ,152.75,2.11e-06,5.22e-10,2.36e-21
J1824-2452C ,240.48,1.30e-06,2.98e-10,2.81e-21
J1824-2452E ,184.53,2.55e-06,4.14e-10,2.54e-21
J1824-2452F ,407.97,6.78e-07,1.30e-10,3.15e-21
J1824-2452G ,169.23,2.93e-06,4.58e-10,2.45e-21
J1824-2452H ,216.01,2.05e-06,3.43e-10,2.70e-21
J1824-2452J ,247.54,2.03e-06,2.87e-10,2.83e-21
J1841+0130 ,33.59,1.10e-04,8.66e-08,6.68e-21
J1843-1113 ,541.81,2.61e-07,7.82e-11,3.24e-21
J1857+0943 ,186.49,4.50e-07,4.09e-10,2.55e-21
J1905+0400 ,264.24,3.36e-07,2.62e-10,2.89e-21
J1909-3744 ,339.32,1.89e-07,1.78e-10,3.06e-21
J1910-5959A ,306.17,8.75e-07,2.10e-10,2.99e-21
J1910-5959B ,119.65,2.83e-06,8.47e-10,2.35e-21
J1910-5959C ,189.49,1.29e-06,4.02e-10,2.57e-21
J1910-5959D ,110.68,2.63e-06,1.07e-09,2.44e-21
J1910-5959E ,218.73,1.06e-06,3.38e-10,2.72e-21
J1911+1347 ,216.17,5.70e-07,3.43e-10,2.70e-21
J1911-1114 ,275.81,2.78e-07,2.46e-10,2.92e-21
J1913+1011 ,27.85,2.93e-04,2.01e-07,8.43e-21
J1939+2134 ,641.93,3.65e-07,5.70e-11,3.27e-21
J1955+2908 ,163.05,3.39e-06,4.80e-10,2.41e-21
J2019+2425 ,254.16,3.07e-07,2.77e-10,2.85e-21
J2033+17 ,168.1,8.65e-07,4.62e-10,2.44e-21
J2051-0827 ,221.8,4.65e-07,3.32e-10,2.73e-21
J2124-3358 ,202.79,6.96e-08,3.71e-10,2.64e-21
J2129-5721 ,268.36,5.13e-07,2.56e-10,2.90e-21
J2145-0750 ,62.3,1.17e-06,8.32e-09,3.84e-21
J2229+2643 ,335.82,2.96e-07,1.81e-10,3.05e-21
J2317+1439 ,290.25,4.68e-07,2.28e-10,2.96e-21
J2322+2057 ,207.97,4.78e-07,3.60e-10,2.66e-21
\end{filecontents*}
\pgfplotstabletypeset[
	col sep=comma,
    	string type,
	columns/Pulsar/.style={column name=Pulsar, column type={| l}},
	columns/Frequency (Hz)/.style={column name= $f_{em}\;(Hz)$, column type={| r |}},
	columns/Ellip LIGO/.style={column name=$\epsilon$, column type={c |}},
	columns/delta div alpha/.style={column name= $\delta \Big/\beta$, column type={c |}},
	columns/graviton mass (eV)/.style={column name= $m_{g}\Big/\sqrt{\beta}\;(\frac{eV}{c^{2}})$, column type={c |}},
	every head row/.style={before row=\hline,after row=\hline},
	every last row/.style={after row=\hline},
	]{116GravLimitsTableB.csv}
	
}
\end{center}

\bibliographystyle{iopart-num}
\bibliography{BimetricMassGrav}

%
%
%


\end{document}